\documentclass[preprint]{aastex}

\begin{document}

\title{X-ray Arc Structures in Chandra Images of NGC5128 (Centaurus A)}
\author{M. Karovska, G. Fabbiano, F. Nicastro, M. Elvis, R.P. Kraft, S.S. Murray}
\affil{Harvard-Smithsonian Center for Astrophysics,60 Garden Street, 
Cambridge, MA}

\begin{abstract}
We describe results from our initial study of the diffuse X-ray
emission structures in NGC 5128
(Cen A) using the Chandra X-ray Observatory
observations.
The high-angular resolution $\it Chandra$ images reveal multi-scale X-ray structures with unprecedented detail and clarity. 
The large scale structures suggest complex symmetry, including a component 
possibly associated with the inner radio lobes, and a separate
component 
with an 
orthogonal symmetry
that may be associated with the galaxy as a whole.
We detect
arc-like soft X-ray structures, extending
to $\sim$ 8 kpc in the direction
perpendicular to the jet. 
These arcs appear to be tracing an ellipse, or a ring seen in projection, and are 
enclosed between the dust and HI emitting gas in the
central region of the galaxy,
and the optical and HI shell fragments at a distance of about 8-10
kpc.
The diffuse X-ray and the optical emission in the arcs could
originate in a region of
interaction (possibly a shock) between the 
infalling material from the outer regions of the galaxy, and the 
cool dust and HI emitting material in the center, or an equatorial outflow
resulting from an
outburst of nuclear activity $\sim$10$^{7}$ years ago.

\end{abstract}


\section{Introduction}

NGC5128 (Centaurus A, Cen A) is 
a giant elliptical early-type galaxy containing the nearest (at 3.5~Mpc; 1{\arcmin}
$\sim$ 1 kpc)
radio-bright AGN. A subparsec scale nuclear region at the center of the galaxy is believed to be associated with a supermassive
black hole ($M_{\bf {BH}}=2 \times 10^{8}M_\sun$; Marconi {\it et al.} 2001).
This low luminosity radio galaxy, a prototype of a major class of
active galaxies (Fanaroff-Riley
Class I), has been observed by virtually every possible astronomical technique from ground and 
space ({\it vis}
review by  Israel 1998).
Multi-wavelength observations of Cen A accumulated over many decades
reveal its  
extremely complex structure.
A radio jet with
huge radio lobes extends across the galaxy to several hundreds of kpc
from the nucleus (Clarke {\it et al.} 1992). Dark bands
stretching across the
middle of the
galaxy, seen in optical and near-IR images of Cen A, are probably due to absorption by dust and other cool
material. 
A system of faint
concentric shell fragments and filaments are also detected in
the deep optical images of Cen A (Malin {\it et al.} 1983).
Radio maps obtained in the 21 cm HI line show emission
originating from cool
material in two distinct regions of the
galaxy: in the central region of the galaxy, along the dust lane,  and
from several shells 10-16{\arcmin} away form the center;
Both the dust lane and the optical and radio shell fragments are 
 thought to be remnants of a merger with a smaller spiral
galaxy (Schiminovich
{\it et al.} 1994).

Cen A was first imaged in X-rays with {\it Einstein} (Feigelson
{\it et al.} 1981), leading to the detection of complex X-ray emission
from a bright nucleus, an extended jet, and a diffuse emission. 
The large scale diffuse emission structures suggested a complex geometry, 
including a component
possibly associated with the jet.
Later observations with 
{\it ROSAT} (Dobereiner {\it et al.} 1996; Turner {\it et al.} 1997) led to the
detection of emission from the southern inner jet radio lobe, and a 
population
of point sources.
Cen A was
observed on several occasions with {\it Chandra},
resulting in studies of the X-ray emission from the jet (Kraft {\it
et al} 2000, 2002a); the point
source 
population (Kraft {\it et al.} 2001); and
sub-arcsecond resolution imaging of the nuclear region  (Karovska {\it et al.}  2001).
In this paper we report the results of the analysis of the 
large-scale diffuse
X-ray emission and the resulting detection
of arcminute-scale 
arc-like X-ray structures.

\section{Chandra Observations}

{\it Chandra} produces
sharper images than any other X-ray telescope to date (FWHM$\sim$0.3{\arcsec} on axis).
This high-spatial resolution results from the innovative design of
this 
observatory in particular 
the high-resolution mirror assembly (Van Speybroeck {\it et
al.} 1997), but also
including
the guidance systems, and
the focal plane detectors.
In addition to the photon positions, {\it Chandra} observations provide information about
their number, energy and time of arrival. 
A detailed description of the Chandra observatory and its capabilities 
can be found in Weisskopf {\it et al.} (2002).
The results
presented in this paper are based on archival observations
 made with both imaging detectors;
the Advanced CCD Imaging Spectrometer (ACIS-I), and the High-Resolution
Camera (HRC-I).

The HRC-I observation (ObsID 463) was carried out on 1999 September 10
for 17 ksec as part of the Orbital
Activation and
Calibration phase. The Cen A nucleus was centered on the aim-point of the detector. 
The HRC-I offers the best resolution images, having a pixel size of
0.13{\arcsec} on the sky, smaller then the mirror PSF.
The total field of view is 33{\arcmin}x33{\arcmin}. 
The characteristics of the detector and an 
initial analysis and results from this 
observation
are discussed by Kraft {\it et al.} (2000).
The ACIS-I observation (ObsID 316) was carried out on 1999 December 5 with 35.9 ks
exposure.
ACIS-I contains an array of four frontside illuminated CCDs with a
total field of view of 16{\arcmin}x16{\arcmin}. 
Each pixel on the ACIS-I  instrument
corresponds to 0.492{\arcsec} on the sky. 
The Cen A nucleus was centered on the ACIS-I I3 CCD.
To analyze the data we used
standard CIAO tools.
\footnote{CIAO is the Chandra Interactive Analysis of Observation's data
analyses system 
package (http://cxc.harvard.edu/ciao)} 

We explored the spatial extent of the X-ray sources in the Cen A
images
by applying an adaptive smoothing algorithm ({\it csmooth}) 
\footnote{{\it csmooth} was adapted from
the {\it asmooth} routine (Harald Ebeling 2000, private communication)}
to the previously binned images.
{\it Csmooth} is a powerful tool for smoothing images containing complex structures
at various spatial scales, since it preserves small-scale spatial signatures
and the associated counts.
The smoothing scale is determined by convolving the input image with 
a gaussian kernel of an increasing size (smoothing scale). 
For each smoothing scale, a 
significance is computed at each pixel location by 
comparing the total counts under the kernel to the expected 
background in the same area.  
The user sets the desired significance threshold value.
The pixels for which the significance exceeds 
this threshold are smoothed at the 
current scale. They are then excluded from subsequent smoothing 
using larger scales.

\section{Results and Analysis}

We detected structures 
in the smoothed HRC-I image at scales ranging from few arcseconds to several
 arcminutes.
Figure 1 shows the 16{\arcmin}x16{\arcmin} field of view from the adaptively smoothed
HRC-I image of Cen A. The outer region of the field of view was excluded 
because
of the edge artifacts created by the smoothing process.
The smoothed image shows diffuse X-ray emission
from Cen A having several distinct components:
a small-scale component 
within a radius of a few arcseconds from the nucleus,
a second complex component from a
region within about 3{\arcmin}-4{\arcmin} radius from the nucleus, and
a large-scale component which includes weak 
X-ray diffuse emission
extending to $\sim$10{\arcmin} from the nucleus.
 
The small-scale component may be associated with the 
innermost portion of the jet or the nuclear disk (Karovska {\it et al.} 2001).
The 3{\arcmin}-4{\arcmin}radius component shows complex structures that roughly follow the ridges of the
dust lane. 
The depression in the X-ray emission in the dust-lane regions
is  probably due to absorption by the dust stretching across the
middle of the galaxy (Feigelson {\it et al.} 1981).
The large scale component includes X-ray emission from regions
along the radio jet (discussed in Kraft {\it et al.} 2001) as well as
a weak 
diffuse emission component, first noted and studied by Feigelson at al (1981), 
showing a 
complex symmetry along two axes: parallel and perpendicular to the jet.

In Figure 2 we show a comparison of the spatial distribution
of the
X-ray 
emission (the central
16{\arcmin}x16{\arcmin} region from the smoothed HRC-I image) with
the emission seen at radio wavelengths: the
VLA 21cm radio continuum map 
\footnote{NASA/IPAC Extragalactic Database; http://nedwww.ipac.caltech.edu}
of the inner lobes of the jet
(Condon {\it et al.} 1996), and 21 cm HI line emission map
\footnote{J. van Gorkom
2002, private communication}. This figure demonstrates the complex
interaction between the various components of this active galaxy.
A comparison of the NE radio 
and X-ray jets 
shows near (but not exact) co-alignment of the general geometry
up to about 4{\arcmin} from the nucleus (Kraft {\it et al.} 2002).
Further away the radio jet develops into a plume and 
the diffuse X-ray emission seems to be distributed in the outer
boundaries of the radio continuum emission, 
following the curvature of the inner lobe (Fig.2).
The component of the diffuse X-ray emission 
associated with
the jet extends to the NE to at least 12{\arcmin} from the nucleus, 
almost as far as the HI emission shell.
The diffuse X-ray emission associated with the SW lobe of the radio
jet
is more complicated, and
extends up to about 10{\arcmin} from the nucleus, reaching to
the SW HI emission shell.
A detailed analysis of the diffuse emission associated with the jet will be discussed
elsewhere (Karovska {\it et al.}, 2002).

In the following we concentrate on 
a distinct component of the diffuse X-ray emission with a symmetry 
orthogonal to the jet. 
The global extent of this emission is similar to
the diffuse emission observed with {\it Einstein} and {\it ROSAT}
(eg. Feigelson {\it et al.} 1981; Fabbiano {\it et al.} 1992; Dobereiner {\it et al.} 1996), but the higher spatial resolution of
Chandra, and the adaptive smoothing of the images, make the detailed morphology of the extended diffuse emission
more apparent.
The most striking part of this large-scale emission are a pair of
diffuse arc-like structures (Figure 1 {\bf A},{\bf B})
extending to ${\sim}$ 8{\arcmin} to the NW, and ${\sim}$7{\arcmin} to
the SE (${\sim}$8 kpc and ${\sim}$7 kpc respectively). They appear 
symmetric about an axis perpendicular to
the direction of the jet.

The NW X-ray arc ({\bf A}) appears to fill a region
enclosed between the inner  HI emission (central 4{\arcmin} region),
and the NW optical and HI shells at a distance of about 8-10{\arcmin} from the center
(Fig.2; see also Fig 1b in Schiminovich
{\it et al.} 1994).
A similar, somewhat fainter X-ray arc is seen to the SE ({\bf B}).This
arc is 
similarly located between the
inner HI emission and the outer optical shells in the SE (Fig 1b in
Schiminovich {\it et al.} 1994). 
The thickness of these arcs is about 2{\arcmin} (${\sim}$ 2 kpc
at the distance of Cen A).
The arc-like structures trace an
ellipse with a
semi-major axis of about 8{\arcmin},
and a semi-minor of about 3{\arcmin}-4{\arcmin}. 
The geometry is consistent with a ring or torus-like
structure seen in a projection (inclination of 60-70${\arcdeg}$ from the
plane of the sky), with an axis of symmetry along the direction of the
inner jet.
As shown in Figure 2,
the inner HI emitting region appears to ``break through'' the X-ray ring
roughly in the E-W direction.
The apparent gaps in the X-ray ring, and the lower diffuse soft X-ray
emission at the inner boundaries of the arcs,
may be due to photoelectric
absorption by cool HI emitting material in the central regions of the galaxy.

We searched for optical emission that could be associated
with the X-ray arcs in the Digitized Sky Survey (DSS)
\footnote{http://archive.stsci.edu/dss} images.
In order to increase the visibility of the faint and low contrast features,
we applied
edge enhancement methods including unsharp masking 
to deep blue and red DSS images.
In the enhanced images we
detected several previously unseen shell fragments and arc-like
regions (Karovska {\it et al.}, 2002), in addition to the known system of shells
(Malin 1983; Gopal-Krishna and Saripalli, 1984).
In particular,
we also detected an arc-like region in the 
blue DSS images which seems to trace
the outer boundaries of the NW X-ray arc (dashed line in Figure 2).
The optical arc is located between the NW X-ray arc and the NW HI
shell (Figure 2).
As with the X-ray arcs, this region could be
tracing an ellipse with a semi-major axis of 
8{\arcmin}, and a semi-minor axis of 4{\arcmin}.
The Western segment of this region has been previously detected
by Malin {\it et al.} (1993) as shell number 8 (see Fig 1b in Schiminovich
{\it et al.} 1994).
Recent UBVRI imaging of this region suggests a young stellar
origin of the emission, possibly associated with a tidal
stream (Peng {\it et al.} 2001).

The NW arc (A in Figure 1) 
was detected in the
soft band (0.5-2 keV) ACIS-I image as well.
To analyze the spectrum, we extracted data from 
a regin centered on the arc.
Background was extracted from an emission-free rectangular
region to the North of the arc
located on the same CCD-chip. 
The NW arc contains $\sim$ 2500 net counts (after removing the point
sources) which is
about 2.5 times 
the number of soft (0.5-2 keV) counts in the background. The
corresponding number of counts in the region bordering the arc to the
South
is about two times lower, and
the signal to noise in the extracted spectrum is not sufficient for
detailed spectral analysis. 
We grouped the arc-region and the background spectra to contain a minimum of 20 (source + local 
background) counts per bin, and performed spectral fitting
assuming that the emission originates from a thermalized
gas. 

We first fitted the data with a model consisting of optically thin thermal emission (the 
model {\em xsmekal} in {\em Ciao's Sherpa}) absorbed by neutral gas,
and 
allowed the temperature and normalization of the thermal plasma to 
vary while keeping both the metallicity of the gas and the column density of 
the absorber frozen to the solar value and to $N_H=9.6 \times10^{20}$cm$^{-2}$,
respectively. The assumed $N_H$ is a sum of the intrinsic
column density measured from the HI map ($1 \times 10^{20}$cm$^{-2}$; see Figure 1b Schiminovich
{\it et al.}) at the South border of the NW arc, and the Galactic column density ($8.6 \times 10^{20}cm^{-2}$; Dickey and Lockman, 1990). 
This produced a statistically acceptable 
fit ($\chi^2_r(dof) = 1.05(80))$, with a temperature of $kT = 0.64 \pm 0.04$
 keV, and a normalization of $1.27 \times 10^{-4}$ (in units of $(10^{-14} 
/ (4 \pi D^2) \times \int_{V} (n_e n_H) dV$, where D is the source distance, 
in cm, $n_e$ and $n_H$ are the electron and proton densities respectively, 
and V is the volume of the emission region). Figure 3
shows the data 
along with the best fitting thermal model (upper panel), and the residuals 
in $\sigma$ (lower panel).
We estimate 
the number density in the gas of $n_e = 5.7 \times 10^{-3}$ 
cm$^{-3}$, and a luminosity of the emitting gas of 
$4.7 \times 10^{38}$ erg s$^{-1}$. In these calculation we
assumed, $n_e \simeq n_H =$ constant, 
a distance of $D \sim 3.5$ Mpc, and an emitting volume of 
$V \sim20$ kpc$^3$ (from the size of the source extraction
region, and
assuming that the region is $\sim2$ kpc deep).
Assuming that the NW arc is about 1/6 of a
ring seen in a projection, we estimate a total luminosity in the ring of $\sim
2.8\times 10^{39}$erg s$^{-1}$.
We then calculate the internal
energy of the gas in the ring of $\sim1.1 \times 10^{55}$ ergs, 
assuming that the mean energy per
particle in the gas is:$(1/2)*kT$.

We also fitted the data allowing the intrinsic equivalent hydrogen column density 
free to vary.
As a result we obtained a somewhat statistically improved fit
with a temperature of
$kT = 0.40 \pm 0.01$, 
$norm = 5.5 \times 10^{-4}$, and $N_H = 3.9 \pm 0.2 \times 10^{21}$
cm$^{-2}$.
The best fit parameters 
are quite different from those obtained in the case with frozen $N_H$.
Using these parameters we calculate a slightly higher density of
$n_e = 1 \times 10^{-2}$ 
cm$^{-3}$, and an almost unchanged internal energy of the gas in the ring of $\sim
1.2 \times10^{55}$ ergs.
We note that our 
assumptions may well be
too simple: the gas might well have non-constant density, and the
absorbing material could be hot. Furthermore,
fits with an equilibrium thermal model may not be correct,
for example if the emission is a result of the shock.
More detailed spectral analysis using more realistic assumptions 
will be presented elsewhere (Karovska {\it et al.}, 2002).

\section{Discussion}

We have detected multi-scale structures in the Chandra images of the
diffuse X-ray emission from Cen A with spatial scales ranging from few
hundred pc to several kpc.
The large-scale diffuse X-ray structures include a set of arcs
which are
symmetric about an axis perpendicular to
the direction of the jet.
The NW arc appears to be associated with the nearby optical
and HI shells. Its outer boundary is surrounded by an arc of
blue stellar emission, possibly a tidally disrupted
stream (Peng {\it et al.} 2001).
These arc-like regions may be fragments of
a ring, or a $\sim$2 kpc
thick torus,  seen in a projection with an outer radius of $\sim$8 kpc
lying in a plane perpendicular to the radio jet. 
The inner boundary of the putative torus borders with dust and HI emission from cool material in the
central region of the galaxy. The apparent gaps in the torus 
could be due
to absorption of the soft X-ray by the cool gas and dust lying in front of the
X-ray emitting gas. 

The diffuse X-ray emission in
the large-scale X-ray ring may be associated with the kpc-scale rolling torus structure in
the central region of Cen A, suggested by
Barker (1999) based on the kinematics studies of the
material in the dust lane. 
The soft X-ray emission may be originating in a region of interaction
(possibly a shock) between the infalling gas from a remnant of the spiral
galaxy, a tidal-tail,
with the material orbiting in the middle of the
elliptical galaxy (including cool dust and HI emitting material).
However, the apparent symmetry of the ring is hard to reconcile with the tidal
origin. Moreover the alignment of the ring perpendicular to the radio
jet suggests an origin related to the nucleus. Hence, 
another possibility is that the emission is a result of an interaction
between a powerful equatorial outflow (or wind) from the nuclear
region of Cen A, and
the system of stellar shells or the galactic ISM.
One could speculate that a giant eruption, resulting for example from
transient nuclear activity, could have produced a galaxy-sized shock wave propagating in the plane 
perpendicular to the direction of the jet.

In a nuclear outburst scenario, we derive a mean velocity of the
ejecta
of $\sim 600$ km/s, by assuming that the X-ray energy in the torus is equal to the kinetic
energy of the ejecta. 
In the case of uniform expansion, using a value of 8 kpc for the outer radius of the torus,
we estimate that the outburst could have occurred $\sim$$10^{7}$ years ago.
This is consistent with the suggested young age of the stars
in the optical arc (Peng {\it et al.} 2001), and with the estimated age of the current burst of star
 formation in the disk ($\sim10^{7}$ years; {\it vis} Israel, 1998).
Based on the estimated energy in the ring, and the X-ray and bolometric luminosities of
the nucleus
($L_{x}\sim 10^{42}$ erg s$^{-1}$, and $L_{bol}\sim 10^{43}$ erg s$^{-1}$, 
respectively; Israel, 1998; Elvis {\it et al.} 1994), the quasar would take $\sim3
\times10^{4}$ years to radiate the energy in the ring, a
short time compared to the time elapsed since the outburst. 

We note that X-ray arc-like structures have been detected in 
other mergers, including Fornax A (Mackie and Fabbiano 1998), and
recently the elliptical galaxy NGC 4636 (Jones et al. 2002)
who also relate them to outbursts of nuclear activity (Ciotti and
Ostriker 1997). 
To determine the nature and the origin of the
emission from the arcs in Cen A and in other similar galaxies,
further multiwavelength observations
are needed
including higher signal to noise X-ray spectra, and detailed studies
of the kinematics.

\acknowledgements

This work was supported by NASA contracts NAS8-39073, The
Chandra Science Center, and the Smithsonian Institution.
We are grateful to Dr. Jacqueline van Gorkom for allowing us to use 
the VLA HI line radio map of Cen A.
The optical images obtained from the DSS archive
were produced at the Space Telescope Science Institute under
U.S. Government grant NAG W-2166. The radio 21cm continuum map was 
obtained from the NASA/IPAC Extragalactic Database.

\newpage

\begin{center}
\includegraphics[width=0.9\textwidth]{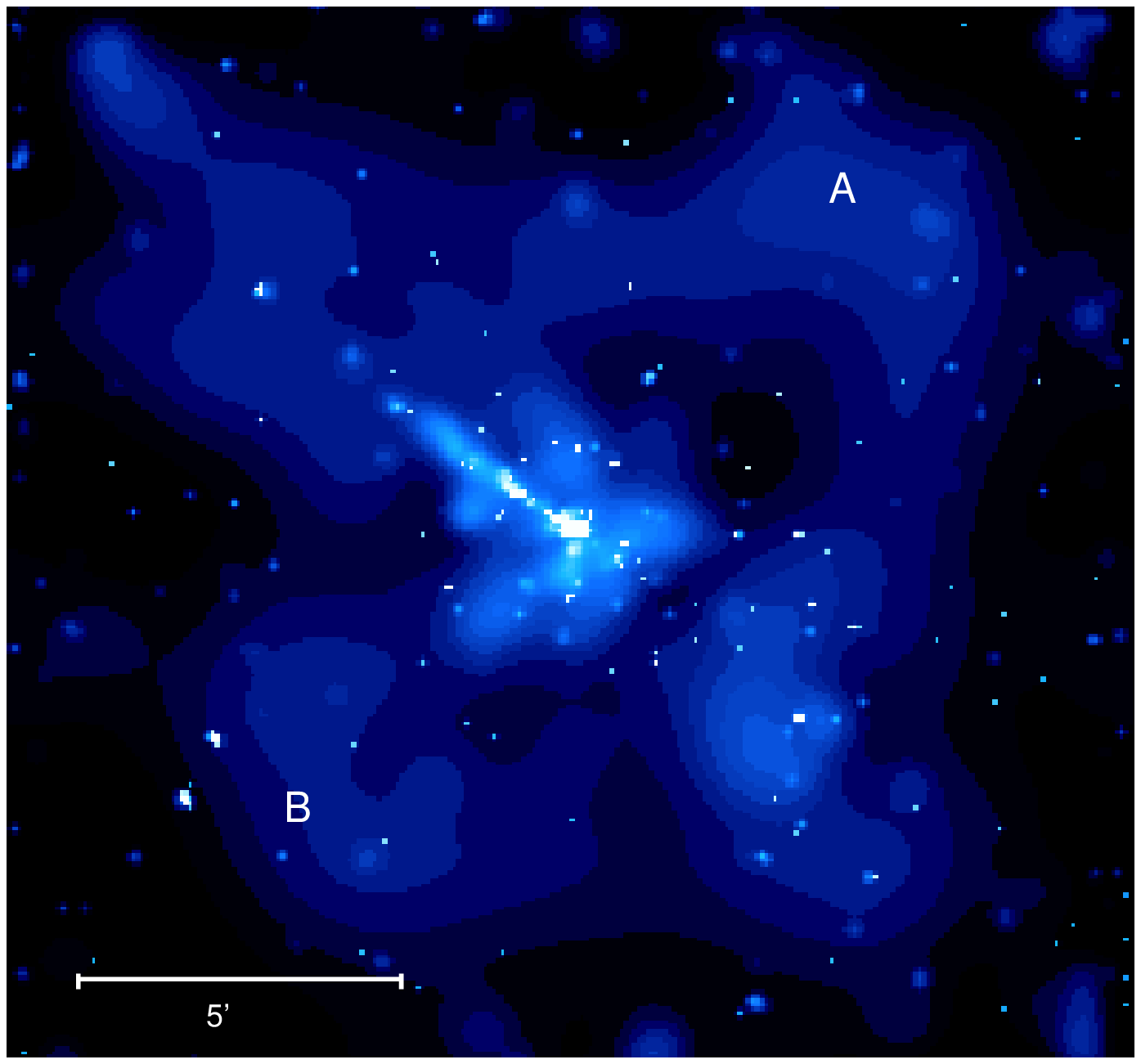}
\end{center}

Figure 1. Adaptively smoothed HRC-I image of the 16{\arcmin}x16{\arcmin} region
centered on the nucleus of Cen A. The pixel size is 4.2{\arcmin}{\arcmin} 
(32 times HRC-I pixel size of 0.13175{\arcmin}{\arcmin}). North is up
and East is to the left. The smoothed image shows simultaneously the 
high-angular resolution bright structures at scales as small
 as few arcseconds and extended faint structures as large as several
arcminutes.

\begin{center}
\includegraphics[width=0.7\textwidth]{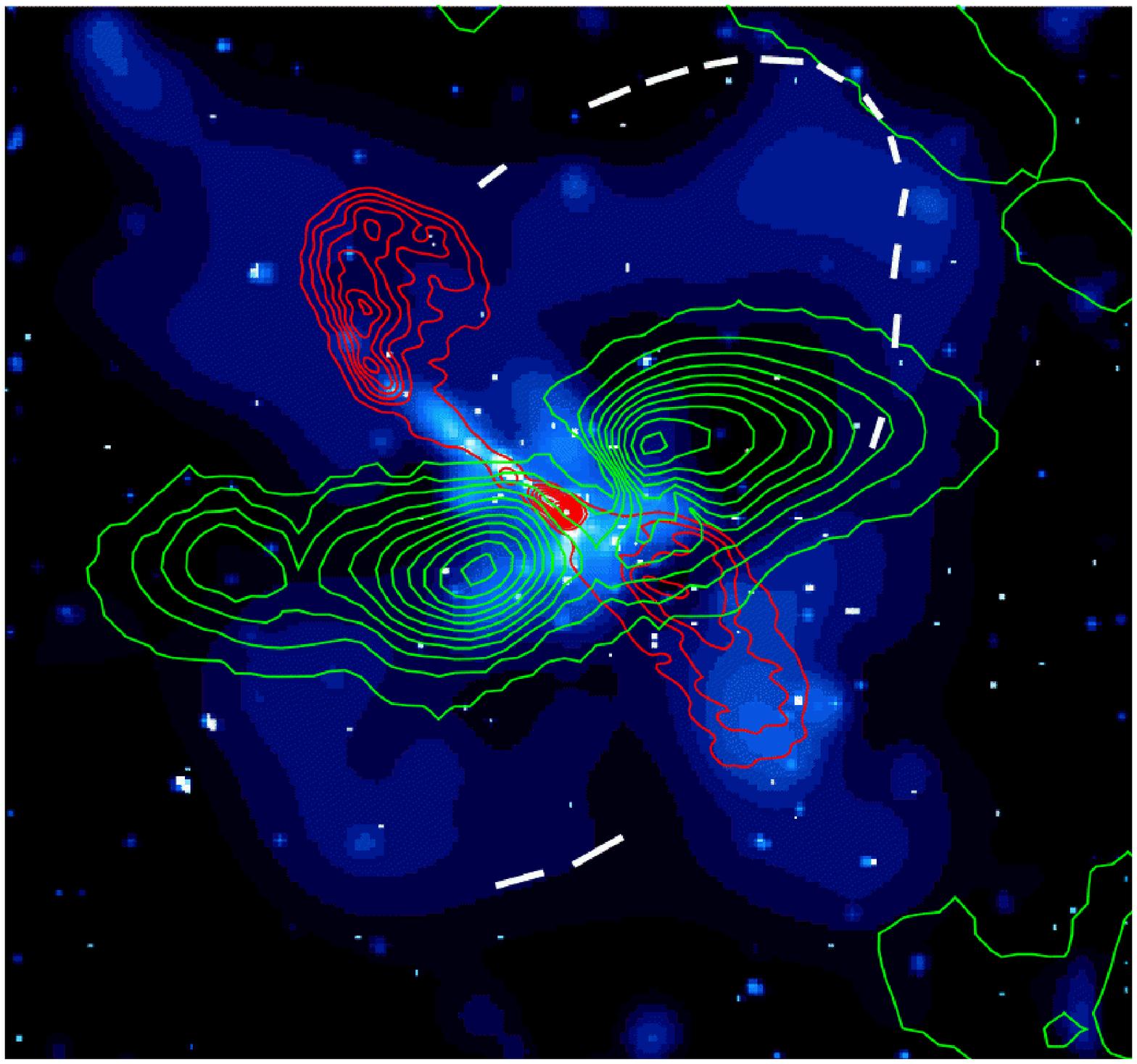}
\end{center}
Figure 2. Cen A HRC-I X-ray image (dark blue) with superimposed the VLA 21 radio continuum map of the
inner jet 
lobes (red contours), and the VLA radio 21 cm line HI gas emission (Schiminovich
{\it et al.} 1994) (green contours).
A white dashed line is a schematical outline of the optical
arc-like shell we detected in the enhanced
blue DSS images.
North is up and East is to the left{}.

\begin{center}
\includegraphics[width=0.9\textwidth]{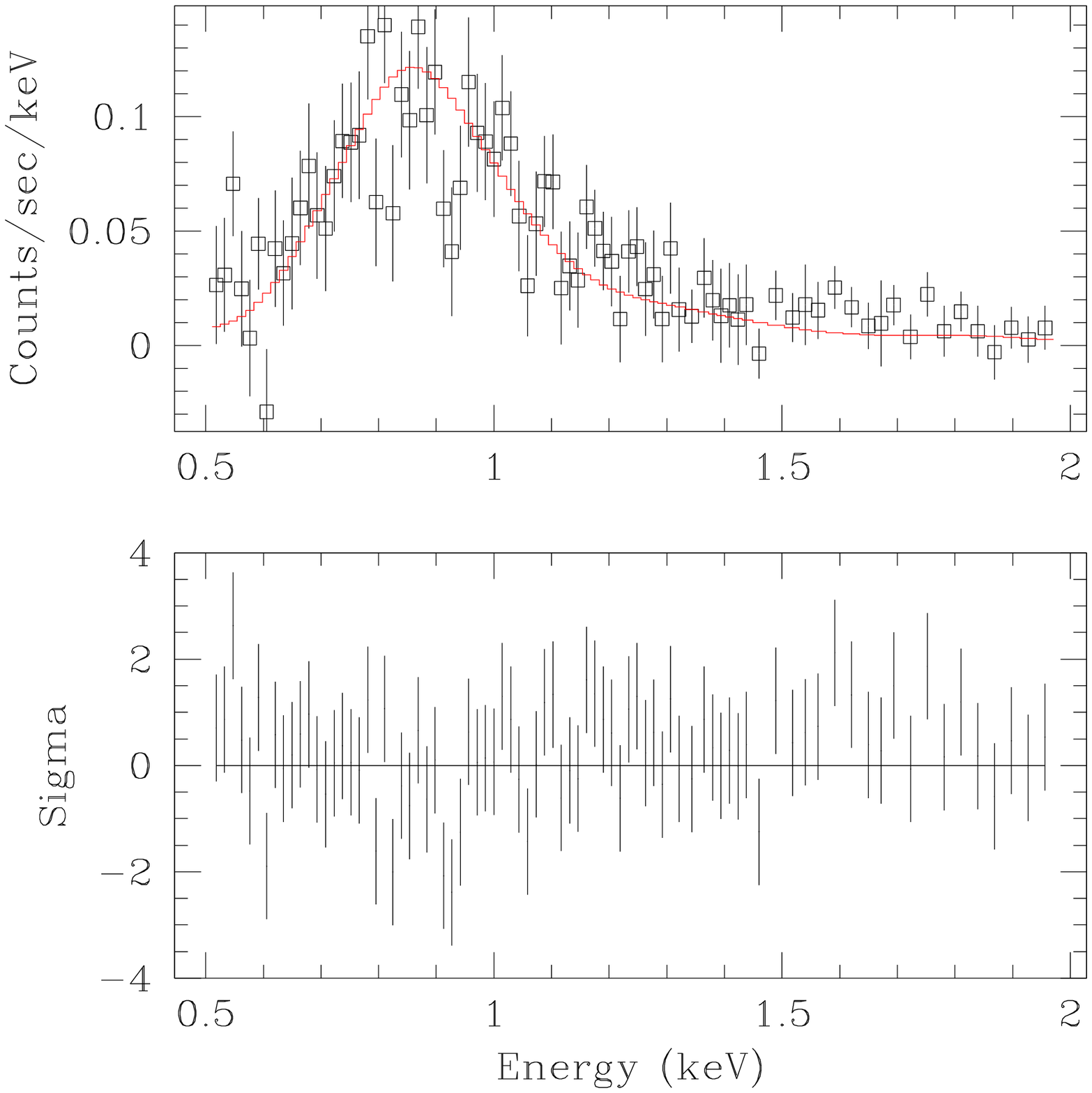}
\end{center}
Figure 3. ACIS spectrum of the NW arc
along with the best fitting thermal model (upper panel), and the residuals 
in $\sigma$ (lower panel).

\end{document}